\documentclass{article}

\PassOptionsToPackage{square,comma,numbers,sort&compress}{natbib}
\usepackage[preprint]{neurips_2025}

\usepackage[utf8]{inputenc} 
\usepackage[T1]{fontenc}    
\usepackage{hyperref}       
\usepackage{url}            
\usepackage{booktabs}       
\usepackage{amsfonts}       
\usepackage{amssymb}        
\usepackage{nicefrac}       
\usepackage{microtype}      
\usepackage{xcolor}         

\usepackage{multirow}
\usepackage{amsmath}
\usepackage{enumitem}
\usepackage{graphicx}
\usepackage{subcaption}
\newcommand{\vis}{\textbf{FramePrompt}}
\title{\vis: In-context Controllable Animation with Zero Structural Changes}

\author{
{\large
Guian Fang },
{\large
Yuchao Gu },
{\large
Mike Zheng Shou$^{\dagger}$}\\
\\
{\large
Show Lab, National University of Singapore}
}

\begin{document}
\maketitle
\let\thefootnote\relax
\footnotemark
\footnotetext{$^{\dagger}$ Corresponding author.}

\begin{abstract}
Generating controllable character animation from a reference image and motion guidance remains a challenging task due to the inherent difficulty of injecting appearance and motion cues into video diffusion models. Prior works often rely on complex architectures, explicit guider modules, or multi-stage processing pipelines, which increase structural overhead and hinder deployment. Inspired by the strong visual context modeling capacity of pre-trained video diffusion transformers, we propose \textbf{FramePrompt}, a minimalist yet powerful framework that treats reference images, skeleton-guided motion, and target video clips as a unified visual sequence. By reformulating animation as a conditional future prediction task, we bypass the need for guider networks and structural modifications. Experiments demonstrate that our method significantly outperforms representative baselines across various evaluation metrics while also simplifying training. Our findings highlight the effectiveness of sequence-level visual conditioning and demonstrate the potential of pre-trained models for controllable animation without architectural changes. %
The project website are available at: \href{https://frameprompt.github.io/}{\color{blue}{\tt Website}}.
\end{abstract}

\section{Introduction}

\begin{figure}
    \centering
    \includegraphics[width=\linewidth]{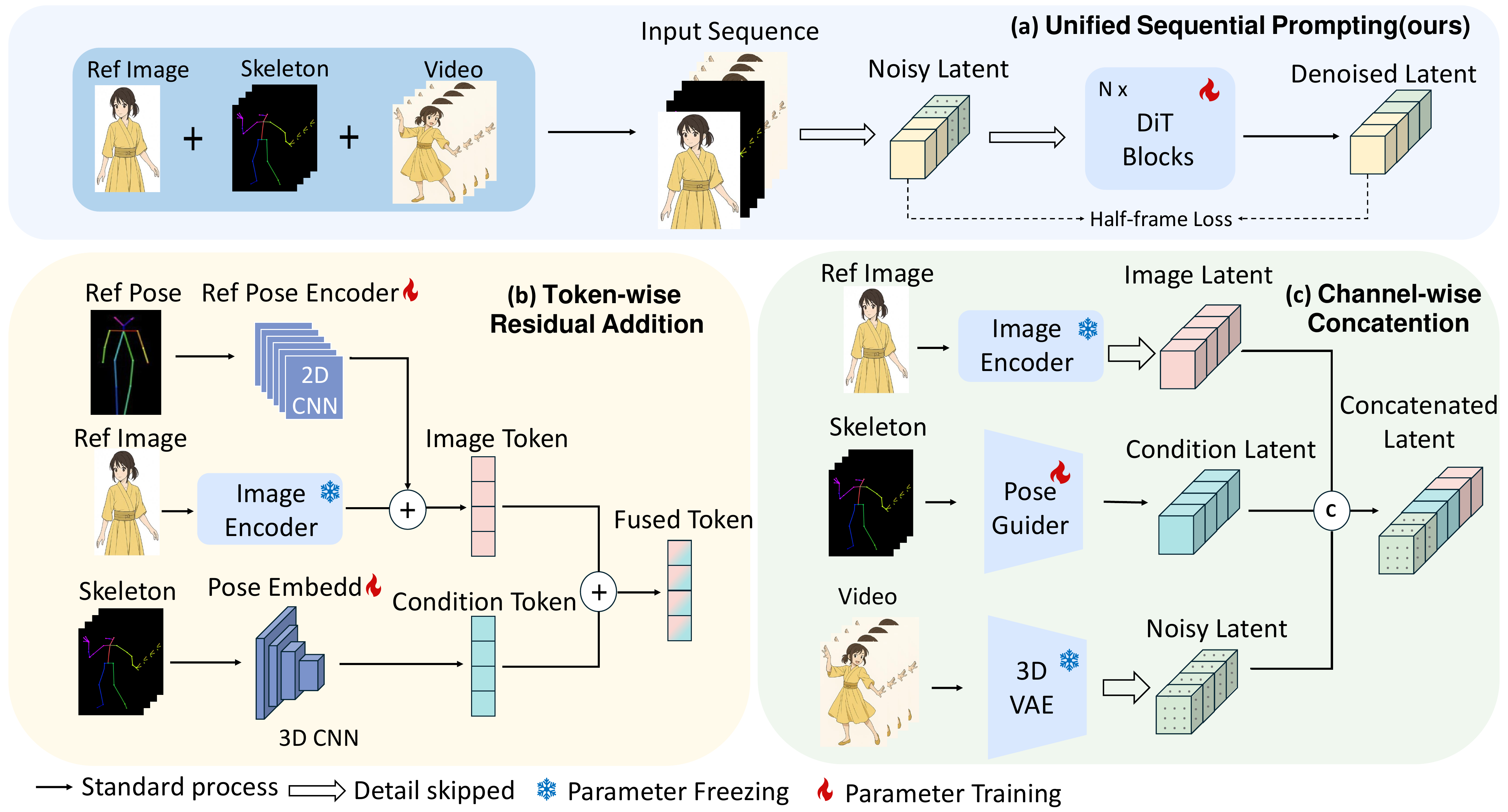}
    \caption{
    \textbf{Comparison of conditioning training strategies in video diffusion models:}
(a) \textbf{Unified Sequential Prompting:} Reference images, skeleton motions, and target frames are concatenated into a single sequence, enabling contextual modeling without architectural changes.
(b) \textit{Token-wise Residual Addition:} Image and pose inputs are embedded and added post-patchification, requiring dedicated encoders. 
(c) \textit{Channel-wise Concatenation:} Latents from conditioning inputs are concatenated along the channel dimension, introducing nontrivial architectural modifications.
    }
    \label{fig:stacked}
\end{figure}

Controllable character animation from a single reference image has become a significant challenge in video generation, with broad applications spanning digital entertainment, virtual avatars, e-commerce, and artistic creation. The primary goal is to synthesize realistic video sequences depicting characters performing specified motions while maintaining visual fidelity to the given reference image. Recent advancements in diffusion models have greatly enhanced the quality of generated visual content, fueling considerable research interest in animation from both academia~\cite{animateanyone, magicanimate, animateanyone2, champ, disco, mimicmotion} and industry~\cite{musepose, mooreaone, realisdance, viggle_ai}.

Previous work on this task often involves treating the source video and target condition differently. For example, U-Net-based diffusion models commonly incorporate specialized modules such as reference networks \cite{animateanyone, magicanimate, animateanyone2, champ} to encode character appearance, pose guidance modules \cite{disco, mimicmotion, realisdance, peng2024controlnext} for motion injection, and temporal modules \cite{musepose, mooreaone} to ensure consistency across frames. Similarly, transformer-based models like \cite{videoxfun, wang2025unianimate, qiu2025skyreels} introduce complex fusion strategies between the source video latent and the corresponding keypoint conditions, along with additional modules. This differentiated handling of source and condition leads to complex, specialized designs and can undermine the original generalization ability of the model.

Inspired by previous large vision models \cite{gemini220250312, gpt4o20250325, liu2025step1x-edit, zhang2025nexus-gen} (LVMs), which organize both the image input and the condition output (e.g., keyposes) within a unified visual sequence, we ask: Can we treat all inputs—the source video, target condition, and target video—as a single visual sequence in a pre-trained video diffusion model? By doing so, we can directly leverage the original context length of a pre-trained video diffusion model for the character animation task without any modifications beyond input sequence organization. However, unlike LVMs, it remains unclear whether pre-trained diffusion models can support heterogeneous input types, as they are trained solely on video data rather than on paired condition and target images, as in LVMs.

In this paper, we make the surprising discovery that pretrained video diffusion transformers inherently possess the contextual capabilities needed for high-quality character animation without requiring architectural changes or additional modules. We, therefore, introduce \vis, a minimalist framework that leverages this property. Specifically, we structure the input sequence into three coherent segments: a reference image, a skeleton-guided motion sequence, and a target video segment. During training, noise addition and denoising loss computations are applied solely to the target segment, effectively reframing the animation task as a future sequence prediction problem based on the provided visual context. In this way, the model learns to extend the given reference and motion cues into coherent, high-quality animations.

Our extensive experiments validate that \vis~achieves highly competitive performance compared to representative methods in the field while being structurally simpler, quicker to train, and easier to deploy. Figure~\ref{fig:stacked} shows the comparison of conditioning training strategies in video DiT. Quantitatively, compared to a representative baseline, \vis~improves SSIM by approximately 5.93\%, PSNR by 20.65\%, LPIPS by 34.95\%, and FVD by 53.87\%. By prioritizing the pre-trained vision context over architectural complexity, \vis~exhibits exceptional robustness across diverse character types and motion patterns.

Our contributions are summarized as follows:
\begin{itemize}
    \item \textbf{Unified Input Sequence Organization}: We propose \vis, a minimalist framework that unifies the reference image, skeleton-guided motion, and target video into a single visual sequence. This reframes character animation as a future sequence prediction task, eliminating the need for specialized modules and significantly simplifying the pipeline.
    
    \item \textbf{Leveraging Pretrained Vision Context}: We demonstrate, for the first time, that pre-trained video diffusion transformers inherently possess the contextual understanding necessary to effectively handle heterogeneous visual input, without any architectural modifications or additional conditioning mechanisms.
    
    \item \textbf{Robustness with Enhanced Efficiency}: Extensive experiments validate that our structurally simpler model significantly outperforms representative methods, demonstrating exceptional robustness across diverse character types, complex motions, and challenging contexts.
\end{itemize}

\section{Related Work}

\subsection{Video Diffusion Models}
Video diffusion models have significantly progressed from initial methods based on U-Net architectures incorporating 3D convolutions, exemplified by pioneering work such as VDM \cite{ho2022video}. These early frameworks primarily relied on spatial-temporal convolutions, which effectively addressed short-range dependencies but struggled with capturing the complex long-range temporal interactions inherent in video data. Recent advancements have led to sophisticated transformer-based architectures, notably the Diffusion Transformers (DiT) \cite{dit, chen2023pixartalpha}, capable of capturing both short- and long-range dependencies through self-attention mechanisms, thereby significantly improving the quality of generated videos.

In parallel, additional improvements have emerged through the incorporation of temporal attention mechanisms \cite{zhou2022magicvideo, wang2023modelscope, guo2023animatediff}, enabling models to dynamically allocate attention across different temporal frames. This further enhances temporal consistency and quality in video outputs. Moreover, multimodal extensions, such as MM-DiT \cite{genmo2024mochi, kong2024hunyuanvideo}, integrate visual data with other modalities, facilitating more flexible and conditionally rich video generation capabilities. Frameworks like Wan-I2V \cite{wan2025} have also introduced diffusion transformers with flexible masking strategies, innovatively reframing video generation tasks as temporal completion problems, thereby pushing forward the boundaries of conditional video synthesis. Open-source initiatives \cite{genmo2024mochi, kong2024hunyuanvideo, HaCohen2024LTXVideo, opensora, lin2024open, jin2024pyramidal, cogvideox} have further democratized these advancements, enabling broader experimentation across diverse video generation tasks.

\subsection{Human Image Animation}
The domain of human image animation has evolved significantly, beginning with Generative Adversarial Networks (GAN)-based frameworks such as the First Order Motion Model \cite{siarohin2019first, siarohin2021motion}, which initially provided foundational techniques for animating static images. Despite their early successes, GAN-based methods faced challenges regarding identity preservation and coherent motion synthesis, prompting the exploration of diffusion-based alternatives such as DisCo \cite{disco} and Animate Anyone \cite{animateanyone}. These diffusion-based frameworks substantially improved identity fidelity and temporal coherence, addressing the limitations of previous methodologies through more controllable training dynamics.

Recent developments have further specialized and refined human image animation methods to address specific challenges within the domain. For example, Animate-X \cite{animatex} effectively handles stylized character animation, RealisDance \cite{realisdance} focuses explicitly on realistic hand motions, and HumanVid \cite{humanvid} optimizes for stable and realistic camera motions. Additionally, transformer-based models \cite{omnihuman, humandit, vace, videoxfun, wang2025unianimate, gan2025humanditposeguideddiffusiontransformer} are emerging as prominent solutions, adeptly handling complex poses and multi-character interactions. These transformer-driven approaches not only enhance animation realism and consistency but also mark a significant shift towards creating more versatile animation models.

\subsection{Visual In-Context Learning}
Visual in-context learning methods have drawn significant inspiration from language models by utilizing the concepts of prompting and in-context examples to guide large vision models (LVMs) \cite{gemini220250312, gpt4o20250325} in interpreting and processing visual data effectively. This paradigm treats visual tasks similarly to textual tasks in large language models, organizing inputs into structured sequences that allow models to leverage contextual relationships for improved task performance. Notable examples include Step1X-Edit \cite{liu2025step1x-edit} and Nexus-Gen \cite{zhang2025nexus-gen}, which demonstrate that structured input sequences can significantly enhance the models’ capability to interpret complex visual scenarios.

Recent studies \cite{tan2024ominicontrol, han2024ace, pu2025lumina} have shown that minimal architectural modifications combined with strategically optimized input organization can profoundly improve visual contextual understanding. These strategies emphasize the role of efficient data structuring over complex architectural enhancements, promoting models that are easier to deploy and maintain. These advancements suggest promising directions for further research, particularly in conditional generation tasks, where coherent contextual modeling can substantially elevate performance outcomes and practical applicability.

\section{Method}

\begin{figure}
    \centering
    \includegraphics[width=\linewidth]{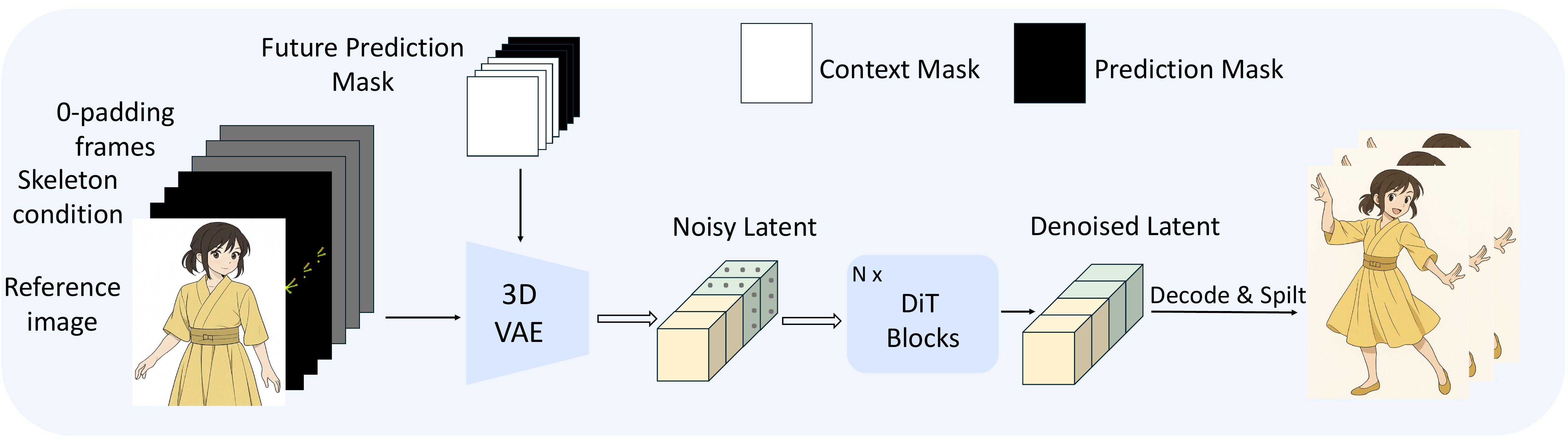}
    \caption{
    \textbf{Inference pipeline of our Unified Sequential Framework:}
    A reference image, a time-ordered skeleton sequence, and zero-padded placeholder frames are concatenated into a single video clip and encoded by a \textbf{3D VAE} to form spatio-temporal latents.
    A binary mask designates \emph{context} tokens (white) to be frozen and \emph{prediction} tokens (black) to be noised.
    Only the prediction tokens undergo iterative denoising through $N$ \textbf{DiT blocks} under a half-frame reconstruction loss.
    The resulting latent is then decoded and split, yielding the future motion frames.
  }
    \label{pipeline}
\end{figure}

Figure~\ref{fig:stacked} provides a schematic comparison of existing conditioning strategies. Our method (Figure~\ref{fig:stacked}a) adopts a unified sequential formulation that simplifies controllable animation by avoiding architectural modifications and specialized encoders. In the following sections, we describe the structured input construction, selective noise diffusion training, and causal inference process that collectively enable high-fidelity motion synthesis with pretrained video diffusion transformers.

\subsection{Preliminary: Image-to-Video Generation}
\label{subsec:preliminary_i2v}

\paragraph{Task Formulation.}
Image-to-Video (I2V) generation synthesizes dynamic video sequences from static reference images, guided by textual prompts or initial frames \cite{zhang2023i2vgen, blattmann2023svd}. We briefly introduce the foundational Wan-I2V framework \cite{wan2025}, which underpins our approach.

\paragraph{Base Model Architecture.}
Recent I2V methods typically extend text-to-video frameworks by concatenating conditional latents with noise latents channel-wise \cite{cogvideox, blattmann2023svd}. Wan-I2V \cite{wan2025} employs a diffusion transformer (DiT) and a masking mechanism to distinguish conditional inputs from outputs. Given a reference image \(I \in \mathbb{R}^{C \times 1 \times H \times W}\), the model appends zero-filled frames to form the guidance sequence, which is encoded by a variational autoencoder (VAE) into latents \(z_c \in \mathbb{R}^{c \times t \times h \times w}\). A binary mask \(M \in \{0,1\}^{1 \times T \times h \times w}\) explicitly differentiates preserved frames from frames to be generated, aligning with latent dimensions via temporal downsampling.

\paragraph{Contextual Enhancement.}
Wan-I2V integrates global contextual features from the CLIP image encoder \cite{radford2021learning}, processes them through an MLP, and injects them via cross-attention mechanisms \cite{ye2023ip}, enhancing visual consistency.

\paragraph{Versatile Masking.}
The masking mechanism supports diverse conditional tasks—such as video continuation, first-last frame transformation, and interpolation—by explicitly delineating preserved frames from generated ones. This approach generalizes complex video tasks as selective temporal completion, forming the conceptual basis for our proposed character animation method.

\subsection{Overview}

To address the complexities and structural limitations inherent in existing approaches to controllable character animation, we propose \textbf{FramePrompt}, a minimalist yet powerful framework. Specifically, we formulate the animation generation task as a future sequence prediction problem, seamlessly integrating diverse visual inputs—reference images, skeletal motions, and target frames—into a unified sequential structure. This approach capitalizes on pretrained video diffusion models without requiring any structural modifications, thereby significantly simplifying both training and inference.

\subsection{Unified Input Sequence and Mask Construction}

Traditional animation models often employ separate modules for different input modalities, complicating both training and deployment. To address this issue, as illustrated in Figure \ref{pipeline}, we organize all inputs into a single coherent visual sequence:

\begin{equation}
X = [I_{\text{ref}}, S_1, \dots, S_T, Z_1, \dots, Z_T] \in \mathbb{R}^{C \times L_{\text{total}} \times H \times W},
\end{equation}

Here, $I_{\text{ref}}$ represents the reference character image, $S_t$ denotes the skeleton frames guiding motion, and $Z_t$ represents the placeholder frames for target video generation. The total length is given by $L_{\text{total}} = 1 + T + T$. During training, these placeholders are initialized with ground truth video frames, whereas during inference, they remain zero-initialized.

We project \( X \) into latent space through a pre-trained Variational Autoencoder (VAE).

\begin{equation}
z = \mathcal{E}(X) \in \mathbb{R}^{c \times L_{\text{total}} \times h \times w}.
\end{equation}

To precisely control the diffusion process, we introduce a binary mask $M \in \{0,1\}^{1 \times L_{\text{total}} \times h \times w}$. This mask delineates between the conditioning and target regions:

\begin{equation}
M_t = \begin{cases}
1, & \text{if } t \leq 1 + T \\
0, & \text{otherwise}
\end{cases},
\end{equation}

thereby enabling selective supervision and noise injection strictly targeting specific frames.

\subsection{Selective Noise Diffusion Training and Inference}

Diffusion-based training typically applies Gaussian noise indiscriminately, which limits precise control over the generated content. To ensure the model specifically focuses on generating novel frames, we selectively apply Gaussian noise only to the target segment:

\begin{equation}
z_t = \sqrt{\alpha_t}\cdot z_0 + \sqrt{1 - \alpha_t}\cdot\epsilon \odot (1 - M),
\end{equation}

Here, \( z_0 \) denotes the clean latent sequence, and \( \epsilon \sim \mathcal{N}(0, I) \).

Accordingly, the model optimizes noise prediction on the target frames using following loss function:

\begin{equation}
\mathcal{L}(\theta) = \mathbb{E}_{z_0, \epsilon, t}\left[\| \epsilon - \epsilon_{\theta}(z_t, t, M)\|^2_2 \cdot (1 - M)\right].
\end{equation}

During inference, we maintain the same input sequence structure but initialize the placeholders with zeros. The latent representations are progressively denoised through iterative diffusion steps:

\begin{equation}
\hat{z}_{\theta}(z_t, t) = z_t - \sqrt{1 - \alpha_t}\cdot\epsilon_{\theta}(z_t, t, M).
\end{equation}

Finally, the VAE decoder reconstructs only the denoised target segment to produce the result:

\begin{equation}
V = \mathcal{D}(\hat{z}_{1+T+1:L_{\text{total}}}),
\end{equation}

where $\mathcal{D}$ denotes the VAE decoder.

\subsection{Implicit Conditioning via Causal Attention}

A crucial advantage of \textbf{FramePrompt} lies in leveraging the causal attention mechanism inherent in pretrained transformer architectures. This mechanism naturally enforces temporal directionality by restricting target frame tokens to attending only to preceding context tokens (reference image and skeleton motion frames). Mathematically, this is defined as:

\begin{equation}
\text{CausalAttention}(Q, K, V) = \text{softmax}\left(\frac{QK^T \odot M}{\sqrt{d_k}}\right)V,
\end{equation}

where $Q, K, V \in \mathbb{R}^{N \times d}$ represent query, key, and value matrices, and mask $M$ explicitly ensures directional information flow.

\subsection{Comparison to Traditional Conditioning Approaches}

Conventional methods, as depicted in Figure \ref{fig:stacked}, employ separate encoders or intricate architectural modifications, relying either on token-wise residual addition or channel-wise concatenation. These approaches inevitably complicate training and limit generalization. In contrast, \textbf{FramePrompt} achieves superior simplicity and efficiency by:

\begin{itemize}
\item Preserving the pretrained model's architectural integrity and generalization.
\item Integrating multiple input modalities into a unified sequence without extra modules.
\item Utilizing the transformer's inherent capability for sequential modeling.
\item Conceptualizing animation explicitly as a context-based future sequence prediction task rather than an explicit conditioning problem.
\end{itemize}

This minimalist design ensures rapid training, straightforward deployment, and robust generalization across diverse characters and complex motion scenarios.

\section{Experiments}

\begin{table}[h]
\footnotesize
\caption{Quantitative comparison with state-of-the-art methods. "Params" denote model size, "Tuning" refers to the fine-tuning strategy, and "Loss" indicates whether the loss is computed on all frames or only the target ones. Best results are in \textbf{bold}, and second-best are \underline{underlined}.}
\label{tab:main_comparison}
\centering
\begin{tabular}{lccccccc}
\toprule
\textbf{Method} & \textbf{Params} & \textbf{Tuning} & \textbf{Loss} & \textbf{SSIM↑} & \textbf{PSNR↑} & \textbf{LPIPS↓} & \textbf{FVD↓} \\
\midrule
MimicMotion†~\cite{mimicmotion2024} & - & - & - & 0.6881 & 12.1742 & 0.3073 & 1038.0991 \\
DisPose†~\cite{li2024dispose} & - & - & - & 0.7821 & 15.3749 & 0.1934 & 576.2598 \\
AnimateAnyone†~\cite{animateanyone,animateanyone2024} & - & - & - & 0.7803 & 15.1771 & 0.2265 & 590.5421 \\
MagicAnimate†~\cite{xu2023magicanimate} & - & - & - & 0.6022 & 7.9100  & 0.3587 & 1267.7109 \\
VideoX-Fun~\cite{videoxfun} & 1.3B & Full & - & 0.7070 & 13.8506 & 0.3525 & 1185.8198 \\
VideoX-Fun~\cite{videoxfun} & 14B & LoRA & - & 0.7622 & 14.2177 & 0.2328 & 373.4316 \\
UniAnimate-DiT~\cite{wang2025unianimate} & 1.3B & Full & - & 0.7958 & 15.3826 & 0.1777 & 325.6171 \\
UniAnimate-DiT~\cite{wang2025unianimate} & 14B & LoRA & - & 0.7831 & 14.7451 & 0.2067 & 487.0679 \\
\midrule
\multirow{8}{*}{\vis~(Ours)} 
& 1.3B & Full & half & \textbf{0.8430} & \textbf{18.5587} & \textbf{0.1156} & \textbf{150.2058} \\
& 1.3B & Full & all & \underline{0.8365} & \underline{18.1249} & 0.1270 & 192.7253 \\
& 1.3B & LoRA & half & 0.8316 & 17.9138 & \underline{0.1238} & \underline{180.9990} \\
& 1.3B & LoRA & all & 0.8284 & 17.5025 & 0.1287 & 202.8364 \\
& 14B & Full & half & 0.8268 & 17.0235 & 0.1307 & 210.2507 \\
& 14B & Full & all & 0.7998 & 15.0748 & 0.1854 & 316.1429 \\
& 14B & LoRA & half & 0.8209 & 17.1880 & 0.1331 & 187.3169 \\
& 14B & LoRA & all & 0.8358 & 17.7344 & 0.1211 & 181.6709 \\
\bottomrule
\end{tabular}

\vspace{2mm}
\raggedright
\footnotesize{
†: U-net-based methods tested directly using released model checkpoints, without fine-tuning.}
\end{table}

In this section, we provide comprehensive experimental validation of the effectiveness of our proposed \vis~~approach. We first detail our experimental setup, including datasets, implementation specifics, and evaluation metrics. We then systematically compare \vis~~with representative state-of-the-art baselines across both synthetic and real-world datasets. Finally, we conduct ablation studies to examine the influence of key design choices.

\subsection{Experimental Setup}

\subsubsection{Datasets}

Previous work predominantly relies on proprietary large-scale datasets, complicating fair comparisons even when using common benchmarks. To mitigate this issue, we synthesized a dataset comprising 394 videos generated using open-source Blender models following the OpenPose format. Each video ensures perfect alignment between skeletal motions and character appearances, thus establishing a controlled, consistent evaluation environment. We reserved 30 videos for testing purposes.

Additionally, we evaluated our method using the publicly available TikTok dataset~\cite{Jafarian_2021_CVPR_TikTok}, which is commonly employed in prior research. Results from this dataset are presented in the Appendix.
\subsubsection{Implementation Details}

For fairness and reproducibility, all DiT-based methods employed identical Wan-I2V initializations and training hyperparameters. Experiments were conducted using two model scales: a fully fine-tuned 1.3-billion-parameter model and a LoRA fine-tuned 14-billion-parameter model. Further implementation specifics, including detailed hyperparameters, are provided in the Appendix.

\subsubsection{Baselines}

We compare \vis~~against representative methods that cover diverse conditioning paradigms: \textbf{Channel-wise Concatenation} (VideoX-Fun~~\cite{videoxfun}), \textbf{Token-wise Residual Addition} (UniAnimate-DiT~\cite{wang2025unianimate}), and several prominent U-Net-based diffusion models (MimicMotion~\cite{mimicmotion2024}, DisPose~\cite{li2024dispose}, AnimateAnyone~\cite{animateanyone}, and MagicAnimate~\cite{xu2023magicanimate}).

\subsubsection{Evaluation Metrics}

Consistent with prior work~\cite{animateanyone}, we employ four standard metrics to assess animation quality: Structural Similarity (SSIM↑), Peak Signal-to-Noise Ratio (PSNR↑), Perceptual Similarity (LPIPS↓), and Fréchet Video Distance (FVD↓). Higher SSIM and PSNR scores, or lower LPIPS and FVD scores, indicate superior performance.

\subsection{Comparison with State-of-the-Art Methods}

\begin{figure*}[htbp]
  \centering
  \begin{subfigure}[b]{0.24\textwidth}
    \includegraphics[width=\textwidth]{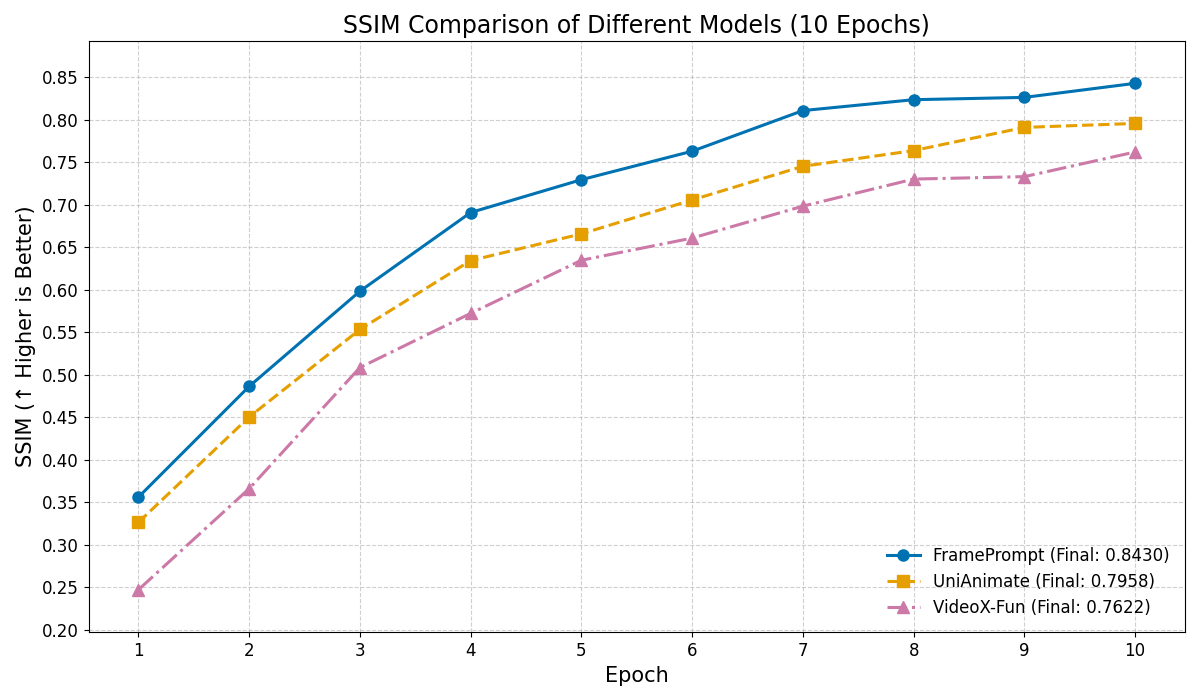}
    \caption{SSIM}
    \label{fig:ssim}
  \end{subfigure}
  \hfill
  \begin{subfigure}[b]{0.24\textwidth}
    \includegraphics[width=\textwidth]{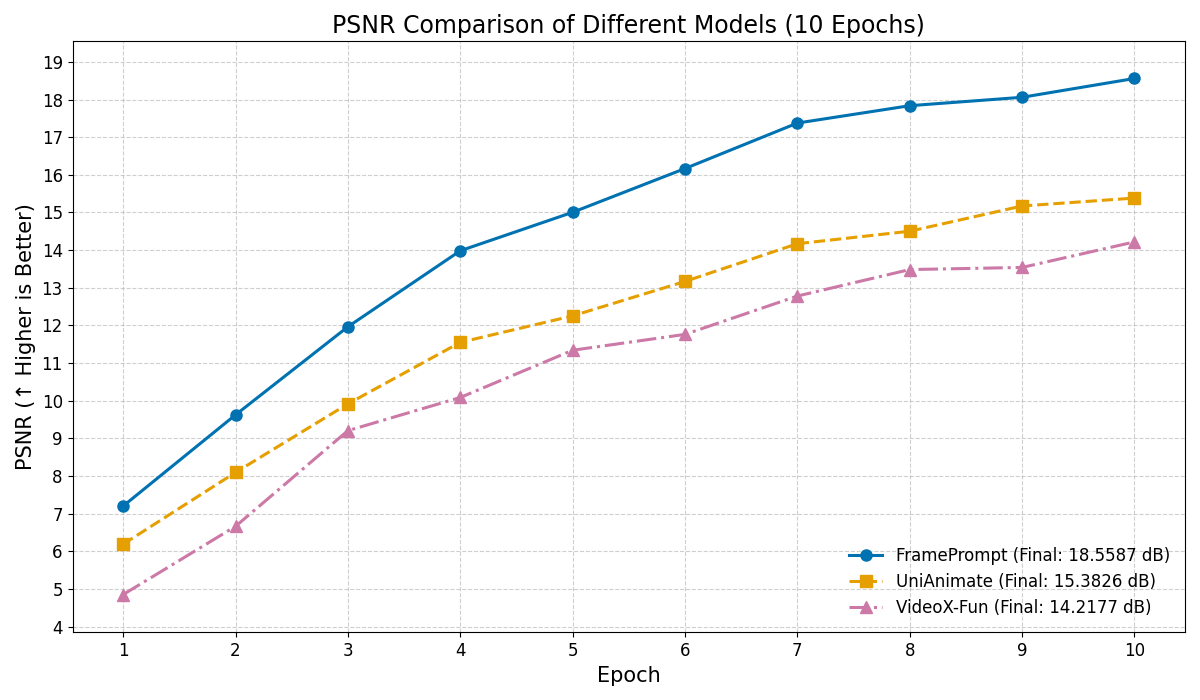}
    \caption{PSNR}
    \label{fig:psnr}
  \end{subfigure}
  \begin{subfigure}[b]{0.24\textwidth}
    \includegraphics[width=\textwidth]{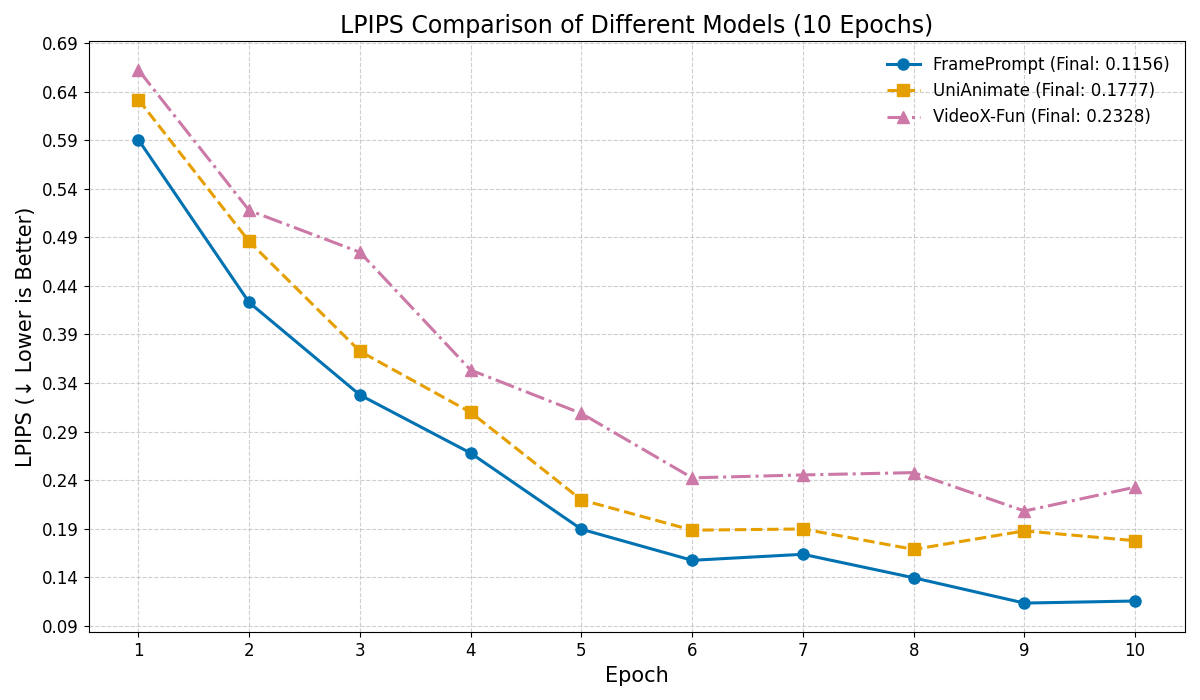}
    \caption{LPIPS}
    \label{fig:lpips}
  \end{subfigure}
  \hfill
  \begin{subfigure}[b]{0.24\textwidth}
    \includegraphics[width=\textwidth]{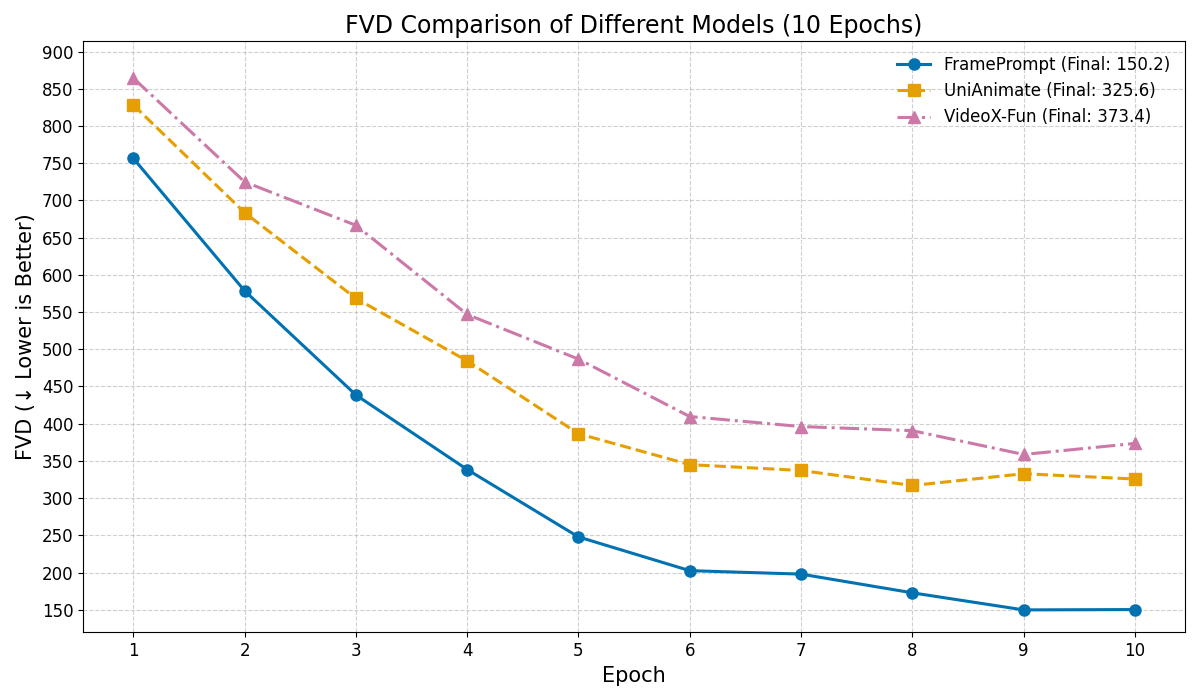}
    \caption{FVD}
    \label{fig:fvd}
  \end{subfigure}
  \caption{Comparison of different animation methods over 10 epochs in terms of SSIM (↑), PSNR (↑), LPIPS (↓), and FVD (↓). FramePrompt consistently outperforms the baselines with faster convergence and better final performance, demonstrating simplified and more effective training.}
  \label{fig:metric_comparison}
\end{figure*}

\paragraph{Quantitative Evaluation.}
Table~\ref{tab:main_comparison} comprehensively summarizes the quantitative performance of our proposed \vis~in comparison with state-of-the-art methods. Under the optimal configuration (1.3B parameters, fully fine-tuned, half-frames loss), \vis~substantially surpasses UniAnimate-DiT (1.3B) by achieving improvements of approximately +5.9\% in SSIM, +20.7\% in PSNR, -34.9\% in LPIPS, and -53.9\% in FVD. These enhancements highlight the effectiveness of our unified sequential input organization in improving perceptual quality and temporal coherence.

\begin{figure}
    \centering
    \includegraphics[width=\linewidth]{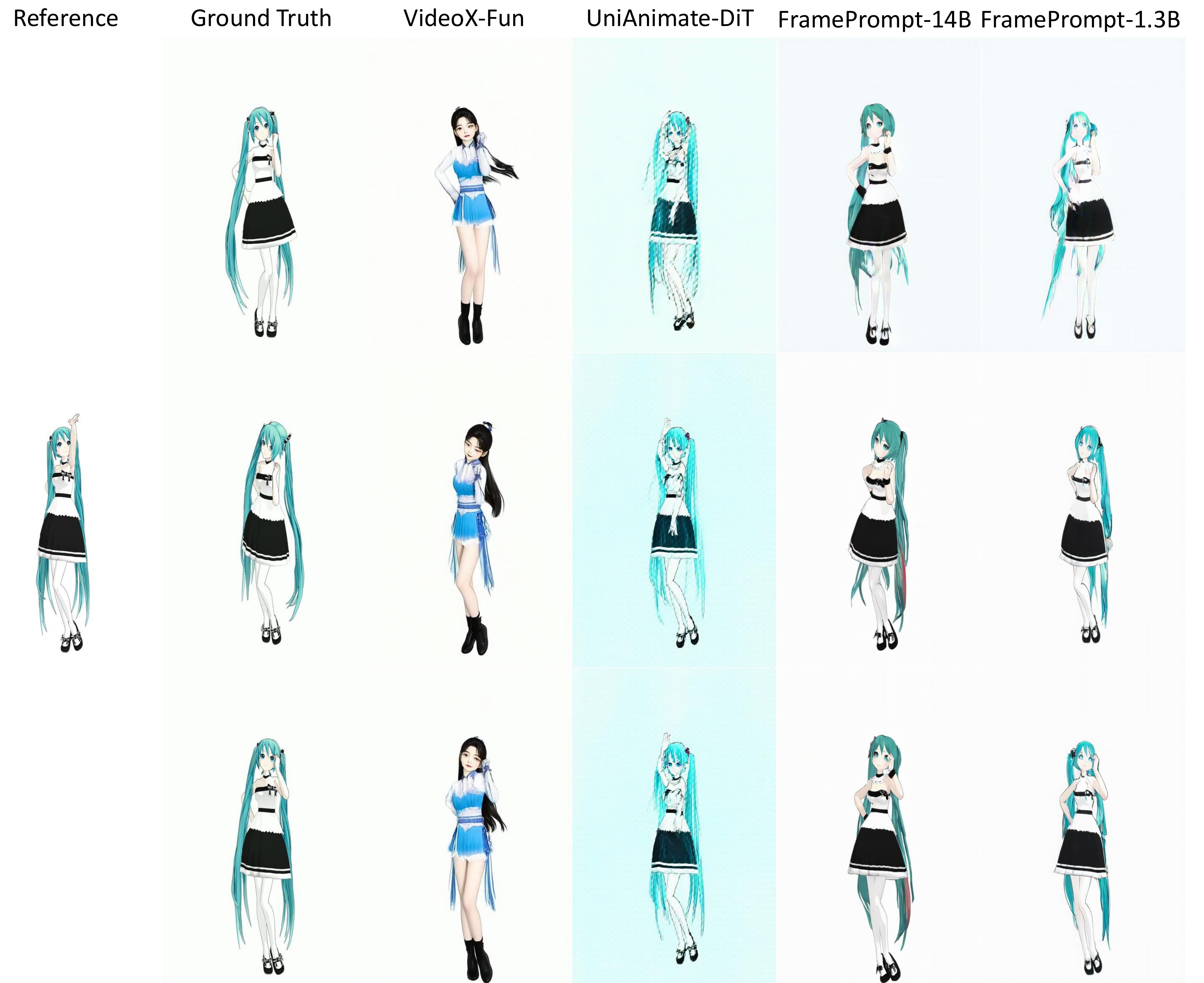}
    \caption{Qualitative comparison of animation results. From left to right: reference image, ground truth, results from VideoX-Fun (14B), UniAnimate-DiT (1.3B), and our \vis~approach (14B and 1.3B), both using half-frames loss under the same settings as the baseline. Additional results are included in the supplementary materials.}
    \label{fig:qualitative_results}
\end{figure}

\paragraph{Qualitative Analysis.}
The qualitative results presented in Figure~\ref{fig:qualitative_results} further substantiate these quantitative advantages. Specifically, VideoX-Fun accurately captures motions but suffers from significant degradation in character identity. Conversely, UniAnimate-DiT preserves character appearance but substantially deviates from the intended motions. \vis~uniquely combines accurate motion capture with identity preservation, particularly excelling in challenging details.

\paragraph{Structural Simplicity and Computational Efficiency.}
A key strength of \vis~lies in its structural simplicity, which markedly contrasts with traditional methods that employ token-wise residual addition or channel-wise concatenation strategies. These conventional methods require dedicated encoders and complex architectural adjustments, leading to increased computational complexity and reduced modularity.

In contrast, our unified sequential prompting approach leverages the inherent contextual modeling capabilities of pretrained video diffusion transformers, eliminating the need for explicit conditioning modules. This significantly reduces model complexity, computational overhead, and memory requirements, thereby enhancing deployability. Specifically, compared to complex baseline architectures, \vis~reduces inference latency and parameter count without sacrificing performance and often improving it.

Figure~\ref{fig:metric_comparison} demonstrates that \vis~achieves faster convergence and superior final performance metrics compared to UniAnimate and VideoX-Fun, underscoring its efficiency and advantages.

\paragraph{Data Efficiency and Generalization.}
Remarkably, \vis~achieves superior or comparable performance using only 394 synthetic videos—a dataset size an order of magnitude smaller than what is typically required by existing methods such as VideoX-Fun and UniAnimate-DiT, which rely on datasets exceeding 10,000 videos. This substantial reduction in training data requirements significantly lowers data acquisition costs and computational demands, highlighting \vis's exceptional data efficiency.

Furthermore, evaluations on a real-world TikTok dataset (see Appendix~\ref{tiktok}) confirm the method's robust generalization capabilities. \vis~maintains performance across diverse and challenging scenarios involving various character identities, complex poses, and cluttered backgrounds, demonstrating practical adaptability and realism.

\subsection{Ablation Studies}

To identify and validate critical design components in \vis, we conduct detailed ablation experiments that cover loss computation strategies, model capacities, and fine-tuning methodologies.

\subsubsection{Impact of Loss Computation Strategy}

To investigate the optimal denoising loss strategy, we compare the application of loss computations on all frames ("all-frames loss") versus only the target frames ("half-frames loss"). Table~\ref{tab:main_comparison} clearly illustrates the superiority of the half-frames loss, significantly enhancing performance metrics. Specifically, the half-frames loss achieves improvements of 0.78\% in SSIM, 2.39\% in PSNR, 8.98\% in LPIPS, and 22.06\% in FVD. This targeted loss strategy proves especially advantageous in scenarios with limited data, enhancing perceptual quality and temporal coherence.

\subsubsection{Influence of Model Capacity}

We further explore how model capacity influences performance by evaluating models with 1.3 billion (1.3B) and 14 billion (14B) parameters. As shown in Table~\ref{tab:main_comparison}, the smaller 1.3B model consistently matches or surpasses the performance of the larger 14B model, notably achieving superior SSIM and significantly lower FVD. This counterintuitive result highlights the importance of optimal model-data alignment, indicating that larger models may underperform when trained on limited datasets due to insufficient generalization.

\subsubsection{Effect of Fine-tuning Methodology}

Finally, we compare full fine-tuning with Low-Rank Adaptation (LoRA) methods. Table~\ref{tab:main_comparison} shows that full fine-tuning consistently yields superior performance across all key metrics compared to LoRA, despite LoRA's higher efficiency. Specifically, full fine-tuning outperforms LoRA in SSIM (0.8430 vs. 0.8316), PSNR (18.56 vs. 17.91), LPIPS (0.1156 vs. 0.1238), and FVD (150.21 vs. 181.00), underscoring the importance of fully adapting global model parameters for detailed structural coherence and enhanced perceptual quality.

\section{Conclusion}

In this paper, we introduced \vis, a minimalist yet highly effective framework for controllable character animation. By reformulating animation as a sequential prediction task, \vis unifies diverse inputs—reference images, skeleton motions, and target video frames—into a single coherent sequence for pretrained video diffusion transformers. This novel approach leverages the inherent contextual understanding of these transformers, thereby eliminating the need for explicit guider networks or architectural modifications and significantly simplifying model design. Our comprehensive experiments demonstrate substantial performance gains over conventional conditioning methods, highlighting the efficacy of prioritizing pretrained visual context over complex architectural additions. Overall, \vis's success suggests broader implications for video generation tasks beyond character animation. Emphasizing pretrained model capabilities and simplified sequence modeling can inspire more efficient and versatile frameworks across various generation challenges.

\newpage
\bibliographystyle{unsrt}
\bibliography{citation}

\begin{thebibliography}{10}

\bibitem{animateanyone}
Li~Hu, Xin Gao, Peng Zhang, Ke~Sun, Bang Zhang, and Liefeng Bo.
\newblock Animate anyone: Consistent and controllable image-to-video synthesis for character animation.
\newblock In {\em {IEEE/CVF} Conference on Computer Vision and Pattern Recognition}, pages 8153--8163, 2024.

\bibitem{magicanimate}
Zhongcong Xu, Jianfeng Zhang, Jun~Hao Liew, Hanshu Yan, Jia{-}Wei Liu, Chenxu Zhang, Jiashi Feng, and Mike~Zheng Shou.
\newblock Magicanimate: Temporally consistent human image animation using diffusion model.
\newblock In {\em {IEEE/CVF} Conference on Computer Vision and Pattern Recognition}, pages 1481--1490, 2024.

\bibitem{animateanyone2}
Li~Hu, Guangyuan Wang, Zhen Shen, Xin Gao, Dechao Meng, Lian Zhuo, Peng Zhang, Bang Zhang, and Liefeng Bo.
\newblock Animate anyone 2: High-fidelity character image animation with environment affordance.
\newblock {\em arXiv preprint arXiv:2502.06145}, 2025.

\bibitem{champ}
Shenhao Zhu, Junming~Leo Chen, Zuozhuo Dai, Zilong Dong, Yinghui Xu, Xun Cao, Yao Yao, Hao Zhu, and Siyu Zhu.
\newblock Champ: Controllable and consistent human image animation with 3d parametric guidance.
\newblock In {\em European Conference on Computer Vision}, volume 15113, pages 145--162, 2024.

\bibitem{disco}
Tan Wang, Linjie Li, Kevin Lin, Yuanhao Zhai, Chung{-}Ching Lin, Zhengyuan Yang, Hanwang Zhang, Zicheng Liu, and Lijuan Wang.
\newblock Disco: Disentangled control for realistic human dance generation.
\newblock In {\em {IEEE/CVF} Conference on Computer Vision and Pattern Recognition}, pages 9326--9336, 2024.

\bibitem{mimicmotion}
Yuang Zhang, Jiaxi Gu, Li-Wen Wang, Han Wang, Junqi Cheng, Yuefeng Zhu, and Fangyuan Zou.
\newblock Mimicmotion: High-quality human motion video generation with confidence-aware pose guidance.
\newblock arXiv preprint arXiv:2406.19680, 2024.

\bibitem{musepose}
Zhengyan Tong, Chao Li, Zhaokang Chen, Bin Wu, and Wenjiang Zhou.
\newblock Musepose: a pose-driven image-to-video framework for virtual human generation.
\newblock {\em arxiv}, 2024.

\bibitem{mooreaone}
Moore-animateanyone.
\newblock {\em https://github.com/MooreThreads/Moore-AnimateAnyone}, 2024.

\bibitem{realisdance}
Jingkai Zhou, Benzhi Wang, Weihua Chen, Jingqi Bai, Dongyang Li, Aixi Zhang, Hao Xu, Mingyang Yang, and Fan Wang.
\newblock Realisdance: Equip controllable character animation with realistic hands.
\newblock {\em arXiv preprint arXiv:2409.06202}, 2024.

\bibitem{viggle_ai}
Viggle ai.
\newblock {\em https://viggleai.io/}, 2024.

\bibitem{peng2024controlnext}
Bohao Peng, Jian Wang, Yuechen Zhang, Wenbo Li, Ming-Chang Yang, and Jiaya Jia.
\newblock Controlnext: Powerful and efficient control for image and video generation.
\newblock {\em arXiv preprint arXiv:2408.06070}, 2024.

\bibitem{videoxfun}
Moore-animateanyone.
\newblock {\em https://github.com/aigc-apps/VideoX-Fun}, 2025.

\bibitem{wang2025unianimate}
Xiang Wang, Shiwei Zhang, Changxin Gao, Jiayu Wang, Xiaoqiang Zhou, Yingya Zhang, Luxin Yan, and Nong Sang.
\newblock Unianimate: Taming unified video diffusion models for consistent human image animation.
\newblock {\em Science China Information Sciences}, 2025.

\bibitem{qiu2025skyreels}
Di~Qiu, Zhengcong Fei, Rui Wang, Jialin Bai, Changqian Yu, Mingyuan Fan, Guibin Chen, and Xiang Wen.
\newblock Skyreels-a1: Expressive portrait animation in video diffusion transformers.
\newblock {\em arXiv preprint arXiv:2502.10841}, 2025.

\bibitem{gemini220250312}
Google Gemini2.
\newblock Experiment with gemini 2.0 flash native image generation, 2025.

\bibitem{gpt4o20250325}
OpenAI.
\newblock Introducing 4o image generation, 2025.

\bibitem{liu2025step1x-edit}
Shiyu Liu, Yucheng Han, Peng Xing, Fukun Yin, Rui Wang, Wei Cheng, Jiaqi Liao, Yingming Wang, Honghao Fu, Chunrui Han, Guopeng Li, Yuang Peng, Quan Sun, Jingwei Wu, Yan Cai, Zheng Ge, Ranchen Ming, Lei Xia, Xianfang Zeng, Yibo Zhu, Binxing Jiao, Xiangyu Zhang, Gang Yu, and Daxin Jiang.
\newblock Step1x-edit: A practical framework for general image editing.
\newblock {\em arXiv preprint arXiv:2504.17761}, 2025.

\bibitem{zhang2025nexus-gen}
Hong Zhang, Zhongjie Duan, Xingjun Wang, Yingda Chen, Yuze Zhao, and Yu~Zhang.
\newblock Nexus-gen: A unified model for image understanding, generation, and editing.
\newblock {\em arXiv preprint arXiv:2504.21356}, 2025.

\bibitem{ho2022video}
Jonathan Ho, Tim Salimans, Alexey Gritsenko, William Chan, Mohammad Norouzi, and David~J Fleet.
\newblock Video diffusion models.
\newblock {\em arXiv preprint arXiv:2204.03458}, 2022.

\bibitem{dit}
William Peebles and Saining Xie.
\newblock Scalable diffusion models with transformers.
\newblock In {\em Int. Conf. Comput. Vis.}, pages 4195--4205, 2023.

\bibitem{chen2023pixartalpha}
Junsong Chen, Jincheng Yu, Chongjian Ge, Lewei Yao, Enze Xie, Yue Wu, Zhongdao Wang, James Kwok, Ping Luo, Huchuan Lu, et~al.
\newblock Pixart-$\alpha$: Fast training of diffusion transformer for photorealistic text-to-image synthesis.
\newblock {\em arXiv preprint arXiv:2310.00426}, 2023.

\bibitem{zhou2022magicvideo}
Daquan Zhou, Weimin Wang, Hanshu Yan, Weiwei Lv, Yizhe Zhu, and Jiashi Feng.
\newblock Magicvideo: Efficient video generation with latent diffusion models.
\newblock {\em arXiv preprint arXiv:2211.11018}, 2022.

\bibitem{wang2023modelscope}
Jiuniu Wang, Hangjie Yuan, Dayou Chen, Yingya Zhang, Xiang Wang, and Shiwei Zhang.
\newblock Modelscope text-to-video technical report.
\newblock {\em arXiv preprint arXiv:2308.06571}, 2023.

\bibitem{guo2023animatediff}
Yuwei Guo, Ceyuan Yang, Anyi Rao, Zhengyang Liang, Yaohui Wang, Yu~Qiao, Maneesh Agrawala, Dahua Lin, and Bo~Dai.
\newblock Animatediff: Animate your personalized text-to-image diffusion models without specific tuning.
\newblock {\em International Conference on Learning Representations}, 2024.

\bibitem{genmo2024mochi}
GenmoTeam.
\newblock Mochi 1.
\newblock \url{https://github.com/genmoai/models}, 2024.

\bibitem{kong2024hunyuanvideo}
Weijie Kong, Qi~Tian, Zijian Zhang, Rox Min, Zuozhuo Dai, Jin Zhou, Jiangfeng Xiong, Xin Li, Bo~Wu, Jianwei Zhang, et~al.
\newblock Hunyuanvideo: A systematic framework for large video generative models.
\newblock {\em arXiv preprint arXiv:2412.03603}, 2024.

\bibitem{wan2025}
Ang Wang, Baole Ai, Bin Wen, Chaojie Mao, Chen-Wei Xie, Di~Chen, Feiwu Yu, Haiming Zhao, Jianxiao Yang, Jianyuan Zeng, Jiayu Wang, Jingfeng Zhang, Jingren Zhou, Jinkai Wang, Jixuan Chen, Kai Zhu, Kang Zhao, Keyu Yan, Lianghua Huang, Mengyang Feng, Ningyi Zhang, Pandeng Li, Pingyu Wu, Ruihang Chu, Ruili Feng, Shiwei Zhang, Siyang Sun, Tao Fang, Tianxing Wang, Tianyi Gui, Tingyu Weng, Tong Shen, Wei Lin, Wei Wang, Wei Wang, Wenmeng Zhou, Wente Wang, Wenting Shen, Wenyuan Yu, Xianzhong Shi, Xiaoming Huang, Xin Xu, Yan Kou, Yangyu Lv, Yifei Li, Yijing Liu, Yiming Wang, Yingya Zhang, Yitong Huang, Yong Li, You Wu, Yu~Liu, Yulin Pan, Yun Zheng, Yuntao Hong, Yupeng Shi, Yutong Feng, Zeyinzi Jiang, Zhen Han, Zhi-Fan Wu, and Ziyu Liu.
\newblock Wan: Open and advanced large-scale video generative models.
\newblock {\em arXiv preprint arXiv:2503.20314}, 2025.

\bibitem{HaCohen2024LTXVideo}
Yoav HaCohen, Nisan Chiprut, Benny Brazowski, Daniel Shalem, Dudu Moshe, Eitan Richardson, Eran Levin, Guy Shiran, Nir Zabari, Ori Gordon, Poriya Panet, Sapir Weissbuch, Victor Kulikov, Yaki Bitterman, Zeev Melumian, and Ofir Bibi.
\newblock Ltx-video: Realtime video latent diffusion.
\newblock {\em arXiv preprint arXiv:2501.00103}, 2024.

\bibitem{opensora}
Zangwei Zheng, Xiangyu Peng, Tianji Yang, Chenhui Shen, Shenggui Li, Hongxin Liu, Yukun Zhou, Tianyi Li, and Yang You.
\newblock Open-sora: Democratizing efficient video production for all, March 2024.

\bibitem{lin2024open}
Bin Lin, Yunyang Ge, Xinhua Cheng, Zongjian Li, Bin Zhu, Shaodong Wang, Xianyi He, Yang Ye, Shenghai Yuan, Liuhan Chen, et~al.
\newblock Open-sora plan: Open-source large video generation model.
\newblock {\em arXiv preprint arXiv:2412.00131}, 2024.

\bibitem{jin2024pyramidal}
Yang Jin, Zhicheng Sun, Ningyuan Li, Kun Xu, Kun Xu, Hao Jiang, Nan Zhuang, Quzhe Huang, Yang Song, Yadong Mu, and Zhouchen Lin.
\newblock Pyramidal flow matching for efficient video generative modeling.
\newblock {\em arXiv preprint arXiv:2410.05954}, 2024.

\bibitem{cogvideox}
Zhuoyi Yang, Jiayan Teng, Wendi Zheng, Ming Ding, Shiyu Huang, Jiazheng Xu, Yuanming Yang, Wenyi Hong, Xiaohan Zhang, Guanyu Feng, Da~Yin, Xiaotao Gu, Yuxuan Zhang, Weihan Wang, Yean Cheng, Ting Liu, Bin Xu, Yuxiao Dong, and Jie Tang.
\newblock {{CogVideoX}}: {{Text-to-Video Diffusion Models}} with {{An Expert Transformer}}.
\newblock In {\em Int. Conf. Learn. Represent.}, 2025.

\bibitem{siarohin2019first}
Aliaksandr Siarohin, St{\'{e}}phane Lathuili{\`{e}}re, Sergey Tulyakov, Elisa Ricci, and Nicu Sebe.
\newblock First order motion model for image animation.
\newblock In {\em Advances in Neural Information Processing Systems}, volume~32, pages 7135--7145, 2019.

\bibitem{siarohin2021motion}
Aliaksandr Siarohin, Oliver~J. Woodford, Jian Ren, Menglei Chai, and Sergey Tulyakov.
\newblock Motion representations for articulated animation.
\newblock In {\em {IEEE/CVF} Conference on Computer Vision and Pattern Recognition}, pages 13653--13662, 2021.

\bibitem{animatex}
Shuai Tan, Biao Gong, Xiang Wang, Shiwei Zhang, Dandan Zheng, Ruobing Zheng, Kecheng Zheng, Jingdong Chen, and Ming Yang.
\newblock Animate-x: Universal character image animation with enhanced motion representation.
\newblock {\em arXiv preprint arXiv:2410.10306}, 2024.

\bibitem{humanvid}
Zhenzhi Wang, Yixuan Li, Yanhong Zeng, Youqing Fang, Yuwei Guo, Wenran Liu, Jing Tan, Kai Chen, Tianfan Xue, Bo~Dai, and Dahua Lin.
\newblock Humanvid: Demystifying training data for camera-controllable human image animation.
\newblock In {\em The Thirty-eight Conference on Neural Information Processing Systems Datasets and Benchmarks Track}, 2024.

\bibitem{omnihuman}
Gaojie Lin, Jianwen Jiang, Jiaqi Yang, Zerong Zheng, and Chao Liang.
\newblock Omnihuman-1: Rethinking the scaling-up of one-stage conditioned human animation models.
\newblock {\em arXiv preprint arXiv:2502.01061}, 2025.

\bibitem{humandit}
Qijun Gan, Yi~Ren, Chen Zhang, Zhenhui Ye, Pan Xie, Xiang Yin, Zehuan Yuan, Bingyue Peng, and Jianke Zhu.
\newblock Humandit: Pose-guided diffusion transformer for long-form human motion video generation.
\newblock {\em arXiv preprint arXiv:2502.04847}, 2025.

\bibitem{vace}
Zeyinzi Jiang, Zhen Han, Chaojie Mao, Jingfeng Zhang, Yulin Pan, and Yu~Liu.
\newblock Vace: All-in-one video creation and editing.
\newblock {\em arXiv preprint arXiv:2503.07598}, 2025.

\bibitem{gan2025humanditposeguideddiffusiontransformer}
Qijun Gan, Yi~Ren, Chen Zhang, Zhenhui Ye, Pan Xie, Xiang Yin, Zehuan Yuan, Bingyue Peng, and Jianke Zhu.
\newblock Humandit: Pose-guided diffusion transformer for long-form human motion video generation, 2025.

\bibitem{tan2024ominicontrol}
Zhenxiong Tan, Songhua Liu, Xingyi Yang, Qiaochu Xue, and Xinchao Wang.
\newblock Ominicontrol: Minimal and universal control for diffusion transformer.
\newblock {\em arXiv preprint arXiv:2411.15098}, 2024.

\bibitem{han2024ace}
Zhen Han, Zeyinzi Jiang, Yulin Pan, Jingfeng Zhang, Chaojie Mao, Chenwei Xie, Yu~Liu, and Jingren Zhou.
\newblock Ace: All-round creator and editor following instructions via diffusion transformer.
\newblock {\em arXiv preprint arXiv:2410.00086}, 2024.

\bibitem{pu2025lumina}
Yuandong Pu, Le~Zhuo, Kaiwen Zhu, Liangbin Xie, Wenlong Zhang, Xiangyu Chen, Pneg Gao, Yu~Qiao, Chao Dong, and Yihao Liu.
\newblock Lumina-omnilv: A unified multimodal framework for general low-level vision.
\newblock {\em arXiv preprint arXiv:2504.04903}, 2025.

\bibitem{zhang2023i2vgen}
Shiwei Zhang, Jiayu Wang, Yingya Zhang, Kang Zhao, Hangjie Yuan, Zhiwu Qin, Xiang Wang, Deli Zhao, and Jingren Zhou.
\newblock I2vgen-xl: High-quality image-to-video synthesis via cascaded diffusion models.
\newblock {\em arXiv preprint arXiv:2311.04145}, 2023.

\bibitem{blattmann2023svd}
Andreas Blattmann, Tim Dockhorn, Sumith Kulal, Daniel Mendelevitch, Maciej Kilian, Dominik Lorenz, Yam Levi, Zion English, Vikram Voleti, Adam Letts, et~al.
\newblock Stable video diffusion: Scaling latent video diffusion models to large datasets.
\newblock {\em arXiv preprint arXiv:2311.15127}, 2023.

\bibitem{radford2021learning}
Alec Radford, Jong~Wook Kim, Chris Hallacy, Aditya Ramesh, Gabriel Goh, Sandhini Agarwal, Girish Sastry, Amanda Askell, Pamela Mishkin, Jack Clark, et~al.
\newblock Learning transferable visual models from natural language supervision.
\newblock In {\em Int. Conf. Mach. Learn.}, pages 8748--8763. PMLR, 2021.

\bibitem{ye2023ip}
Hu~Ye, Jun Zhang, Sibo Liu, Xiao Han, and Wei Yang.
\newblock Ip-adapter: Text compatible image prompt adapter for text-to-image diffusion models.
\newblock {\em arXiv preprint arXiv:2308.06721}, 2023.

\bibitem{mimicmotion2024}
Yuang Zhang, Jiaxi Gu, Li-Wen Wang, Han Wang, Junqi Cheng, Yuefeng Zhu, and Fangyuan Zou.
\newblock Mimicmotion: High-quality human motion video generation with confidence-aware pose guidance.
\newblock {\em arXiv preprint arXiv:2406.19680}, 2024.

\bibitem{li2024dispose}
Hongxiang Li, Yaowei Li, Yuhang Yang, Junjie Cao, Zhihong Zhu, Xuxin Cheng, and Chen Long.
\newblock Dispose: Disentangling pose guidance for controllable human image animation.
\newblock {\em arXiv preprint arXiv:2412.09349}, 2024.

\bibitem{animateanyone2024}
Moore Threads.
\newblock Animateanyone: Consistent human motion transfer in realistic videos.
\newblock \url{https://github.com/MooreThreads/Moore-AnimateAnyone}, 2024.
\newblock Accessed: 2025-05-15.

\bibitem{xu2023magicanimate}
Zhongcong Xu, Jianfeng Zhang, Jun~Hao Liew, Hanshu Yan, Jia-Wei Liu, Chenxu Zhang, Jiashi Feng, and Mike~Zheng Shou.
\newblock Magicanimate: Temporally consistent human image animation using diffusion model.
\newblock In {\em arXiv}, 2023.

\bibitem{Jafarian_2021_CVPR_TikTok}
Yasamin Jafarian and Hyun~Soo Park.
\newblock Learning high fidelity depths of dressed humans by watching social media dance videos.
\newblock In {\em Proceedings of the IEEE/CVF Conference on Computer Vision and Pattern Recognition (CVPR)}, pages 12753--12762, June 2021.

\end{thebibliography}

\newpage
\appendix
\section{Experimental Details}

\subsection{Implementation Details}
In this section, we provide detailed implementation information for our \vis~framework, including training setup, data processing, and model configuration.
\subsubsection{Training Setup}
Our experiments were conducted on a system equipped with 8 NVIDIA H100-80GB GPUs. All comparative experiments used identical hardware configurations and hyperparameter settings to ensure fair comparison. Our key training parameters are as follows:
\begin{table}[h]
\caption{Training Hyperparameters}
\label{tab:hyperparams}
\centering
\begin{tabular}{lc}
\toprule
\textbf{Parameter} & \textbf{Value} \\
\midrule
Learning Rate & $1 \times 10^{-4}$ \\
Maximum Epochs & 10 \\
Batch Size & 1 \\
Gradient Accumulation Steps & 1 \\
LoRA Rank & 128 \\
LoRA Alpha & 128 \\
Training Strategy & DeepSpeed ZeRO-2 \\
Gradient Checkpointing & Enabled \\
Random Seed & 42 \\
Inference Steps & 50 \\
\bottomrule
\end{tabular}
\end{table}

We employed the AdamW optimizer with a consistent random seed of 42 across all training and generation processes to ensure reproducibility of our results.

\subsubsection{Model Configuration}
We initialized our models with Wan-I2V pretrained weights. For the 14B parameter model, we employed LoRA fine-tuning with targeted modules including query, key, value, output projections, and feed-forward networks. For the 1.3B parameter model, we conducted full fine-tuning. The image dimensions were set to $832 \times 480$ pixels, and we processed 81 frames per sequence.
\begin{align}
\text{LoRA Target Modules} &= \{\text{q, k, v, o, ffn.0, ffn.2}\} \\
\text{Image Dimensions} &= 832 \times 480 \text{ pixels} \\
\text{Sequence Length} &= 81 \text{ frames}
\end{align}
\subsubsection{Data Processing}
To ensure efficient processing of high-resolution images, we implemented tiled encoding in the VAE with the following configuration:
\begin{align}
\text{Tile Size} &= 34 \times 34 \\
\text{Tile Stride} &= 18 \times 16
\end{align}
This tiling approach significantly reduced the VRAM requirements while maintaining high-quality latent representations.

\newpage
\section{More Results}
\subsection{Quantitative Results on Real-World Data}
\label{tiktok}
To address potential domain gaps arising from synthetic datasets, we conducted additional experiments on the TikTok~\cite{Jafarian_2021_CVPR_TikTok} dataset, following the same training/testing split ratio as in previous work. Table \ref{tab:new_dataset_comparison} presents the results of these experiments.

\begin{table}[h]
\caption{Ablation study results on the TikTok dataset.}
\label{tab:new_dataset_comparison}
\centering
\begin{tabular}{lccccccc}
\toprule
\textbf{Method} & \textbf{Params} & \textbf{Tuning} & \textbf{SSIM↑} & \textbf{PSNR↑} & \textbf{LPIPS↓} & \textbf{FVD↓} \\
\midrule
VideoX-Fun~\cite{videoxfun} & 1.3B & Full & 0.4754 & 12.2225 & 0.4048 & 509.8587 \\
VideoX-Fun~\cite{videoxfun} & 14B & LoRA & 0.4162 & 10.6200 & 0.4476 & 568.7819 \\
UniAnimate-DiT~\cite{wang2025unianimate} & 1.3B & Full & \underline{0.5267} & \underline{13.4025} & \underline{0.3821} & \underline{436.1153} \\
UniAnimate-DiT~\cite{wang2025unianimate} & 14B & LoRA & 0.4148 & 12.6016 & 0.5787 & 453.6322 \\
\vis~(Ours) & 1.3B & Full & 0.6237 & 14.3657 & 0.2513 & 394.4333 \\
\vis~(Ours) & 14B & LoRA & \textbf{0.6364} & \textbf{15.1999} & \textbf{0.2220} & \textbf{289.1073} \\
\bottomrule
\end{tabular}
\end{table}

The results on the TikTok dataset demonstrate that our \vis~approach maintains its performance advantage over baseline methods even with real-world data. This indicates that the effectiveness of our unified sequential framework is generalizable across domains, not merely an artifact of synthetic training data. Notably, both our 1.3B and 14B parameter models significantly outperform all baseline methods across all evaluation metrics.

Unlike in the synthetic dataset experiments, the 14B parameter model with LoRA fine-tuning achieves the best performance on the TikTok dataset, with substantial improvements over its closest competitor, UniAnimate-DiT 1.3B: 20.83\% higher SSIM, 13.41\% higher PSNR, 41.90\% lower LPIPS, and 33.71\% lower FVD. This suggests that for more complex real-world scenarios with greater variability in appearance, motion patterns, and backgrounds, the additional capacity of larger models becomes beneficial when properly leveraged through our unified sequential framework.

The consistent performance advantage across both synthetic and real-world datasets validates the robustness of our approach and its effectiveness in capturing the complex relationships between reference appearances and motion sequences in diverse scenarios. These results further support our hypothesis that treating heterogeneous visual inputs within a unified sequential framework enables more effective leverage of pretrained vision context than traditional conditioning approaches, regardless of the data domain.

\subsection{Qualitative Results}

\begin{figure}[h]
    \centering
    \includegraphics[width=\linewidth]{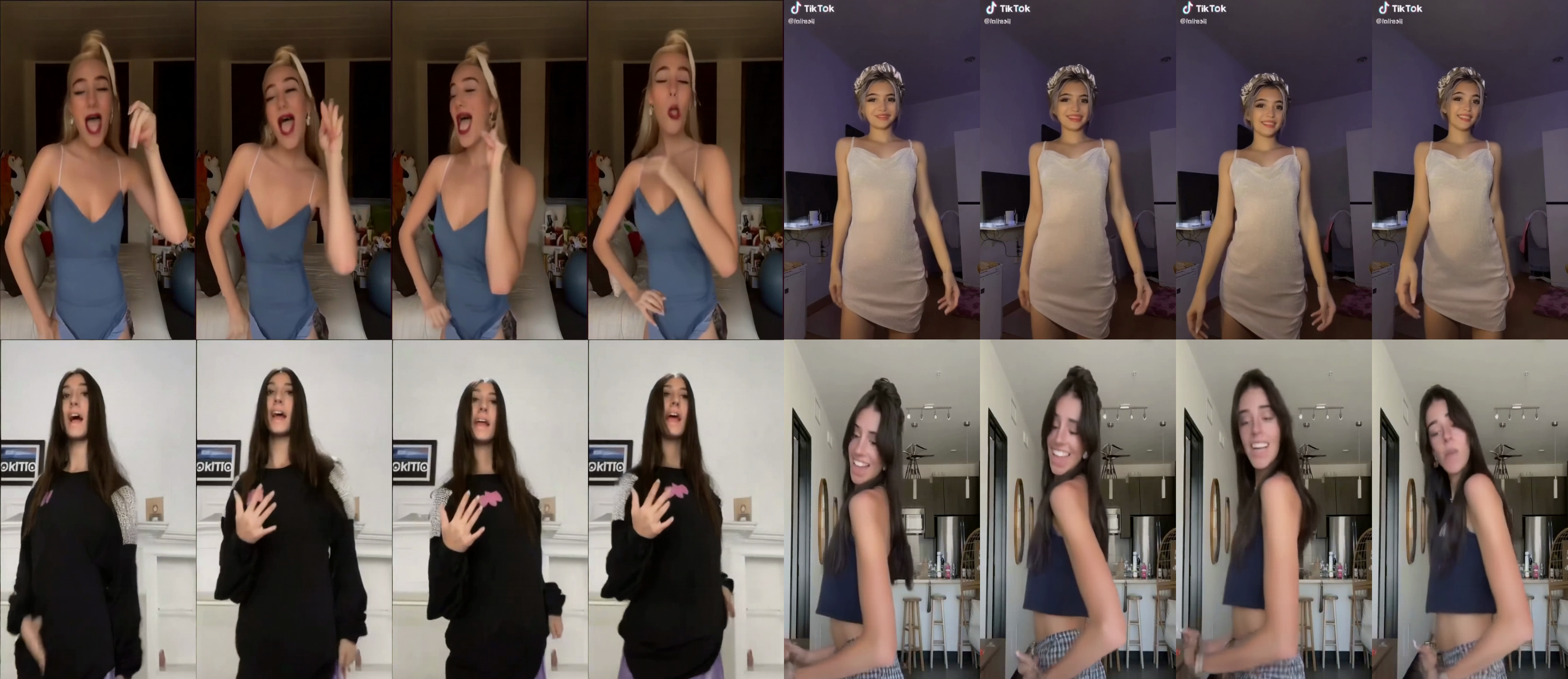}
    \caption{Showcase of our \vis~approach generating diverse animations from the TikTok dataset. Each row displays a different test case, showing sequential frames from the generated animation. Our method successfully captures a wide range of motion patterns while maintaining consistent appearance details throughout the sequence.}
    \label{fig:tiktok_showcase}
\end{figure}

\begin{figure}[h]
    \centering
    \includegraphics[width=\linewidth]{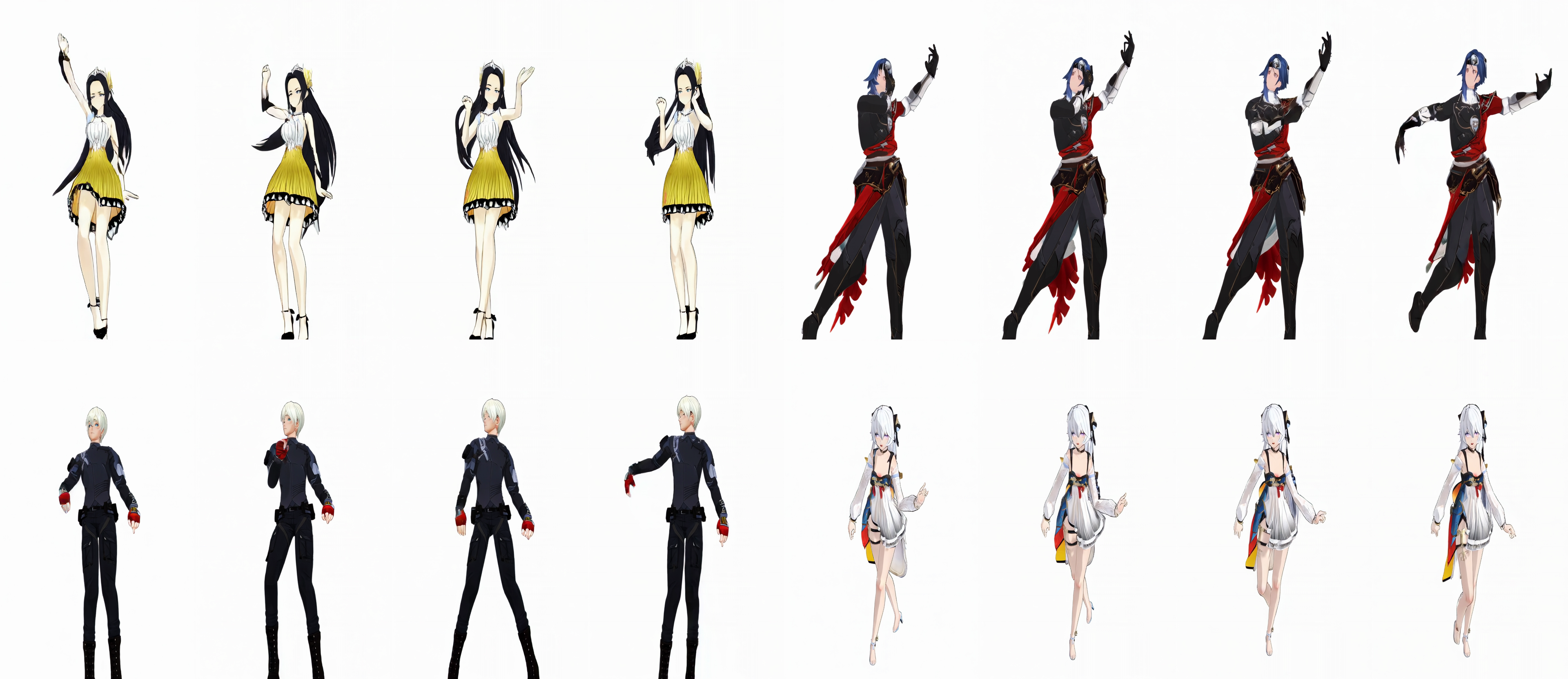}
    \caption{Examples of animations generated by our \vis~approach on the synthetic dataset. Each row demonstrates our model's ability to faithfully reproduce complex character appearances and motion patterns while maintaining temporal consistency across frames.}
    \label{fig:synthetic_showcase}
\end{figure}

In addition to the quantitative metrics, we provide visual examples to qualitatively assess the performance of our \vis~approach across both synthetic and real-world datasets. Figures \ref{fig:tiktok_showcase} and \ref{fig:synthetic_showcase} display representative examples from our model's generated animations.

As shown in Figure \ref{fig:tiktok_showcase}, our approach generates high-quality animations across diverse scenarios from the real-world TikTok dataset. The examples demonstrate our model's capability to handle various types of human motions, from subtle expressions and gestures to more dynamic movements, all while maintaining consistent appearance details from the reference images.

Similarly, Figure \ref{fig:synthetic_showcase} illustrates our method's effectiveness on the synthetic dataset. These examples highlight our model's ability to capture precise character appearances and complex motion patterns in controlled environments, further validating the generalizability of our unified sequential framework across different data domains.

Several key strengths of our approach are evident in these visual results:

\begin{enumerate}
    \item \textbf{Motion Diversity}: Our model effectively handles a wide range of motion patterns, from expressive hand gestures and facial movements to full-body dancing motions, demonstrating its adaptability to different animation scenarios.
    
    \item \textbf{Temporal Consistency}: Across all examples, our method maintains smooth transitions between frames with minimal flickering or jitter, producing natural-looking animations that preserve the temporal coherence of human movements.
    
    \item \textbf{Identity Preservation}: The character appearance remains consistent throughout the animation sequences, with accurate preservation of facial features, hairstyles, clothing details, and other distinctive attributes from the reference images.
    
    \item \textbf{Cross-Domain Performance}: The consistent quality across both synthetic and real-world examples demonstrates our method's robustness to domain shifts and its effectiveness in diverse visual contexts.
\end{enumerate}

These visual results complement our quantitative findings, providing concrete evidence of our method's effectiveness in generating high-quality character animations. By reformulating character animation as a future sequence prediction problem within a unified visual context, our approach demonstrates superior ability to capture the complex relationships between reference appearance and motion guidance, translating these into coherent and visually appealing animations across both controlled synthetic environments and challenging real-world scenarios.
\newpage
\section{Broader Impacts}

Our \vis~framework has several potential societal implications. On the positive side, it democratizes character animation by reducing technical complexity and computational requirements, enabling broader creative expression across small studios, independent creators, and educational institutions. This could accelerate content creation in entertainment, education, and interactive media.

However, we acknowledge potential concerns including: (1) the possibility of unauthorized synthetic character animations, (2) intellectual property considerations regarding character appearances, and (3) potential labor market transitions for traditional animators. 

To address these concerns, we recommend: (1) implementing digital watermarking for synthetic content identification, (2) developing clear usage guidelines that respect intellectual property rights, and (3) supporting initiatives to help traditional animation professionals transition to complementary roles in AI-assisted workflows.

We believe the research community should continue working on balancing technical innovation with ethical considerations, developing frameworks that maximize creative possibilities while establishing appropriate safeguards against misuse.

\section{Limitations}

While our \vis~framework demonstrates strong performance for character animation, we acknowledge several limitations:

\paragraph{Boundary Frame Quality.} We observe lower visual quality in the first and last frames of generated sequences compared to middle frames. This is likely due to diffusion models' inherent challenges with handling abrupt transitions between conditioning and target sequences. This limitation affects applications requiring seamless looping or concatenated animations.

\paragraph{Sequence Length Constraints.} Our approach is constrained by the maximum context length of pretrained video diffusion transformers, limiting animation duration in a single generation pass. This is particularly relevant for our sequential framework as both conditioning inputs and generation outputs consume context tokens. While generating segments and concatenating them could address longer sequences, this may exacerbate the boundary quality issues mentioned above.

These limitations present promising directions for future research, including developing techniques for improving boundary frame quality and extending effective context length while maintaining our framework's architectural simplicity.


\end{document}